\documentclass[twocolumn,showpacs,superscriptaddress,prl,aps]{revtex4}
\usepackage{amssymb}
\usepackage{amsmath}
\usepackage{amsfonts}
\usepackage{graphics}
\usepackage{epic}
\usepackage{eepic}
\usepackage{color}
\usepackage{epsfig}
\usepackage{latexsym}

\begin{document}

\def\ra{\rangle}
\def\la{\langle}
\def\be{\begin{equation}}
\def\ee{\end{equation}}
\def\nn{\nonumber}

\newcommand{\bra}[1]{\left\langle #1 \right\vert}
\newcommand{\ket}[1]{\left\vert #1 \right\rangle}

\title{Spectral Effects of Fast Response Cross Kerr Non-Linearity on Quantum Gate}

\author{Patrick M. Leung}
\author{Timothy C. Ralph}
\affiliation{Centre for Quantum Computer Technology, Department of
Physics,  University of Queensland, Brisbane 4072, Australia}

\author{William J. Munro}
\author{Kae Nemoto}
\affiliation{Hewlett-Packard Laboratories, Filton Road, Stoke Giord,
Bristol BS34 8QZ, United Kingdom}
\affiliation{National Institute of Informatics, 2-1-2 Hitotsubashi,
Chiyoda-ku, Tokyo 101-8430, Japan}

\email{pmleung@physics.uq.edu.au}

\date{\today}

\pacs{03.67.Lx, 42.50.-p}

\begin{abstract}

According to idealized models, a strong Kerr non-linearity may be used to build
optical quantum gates for optical quantum information processing by
inducing conditional phase shifts on quantum states. Recently, Shapiro (PRA 73,
062305 (2006)) argued that for a Kerr medium with non-instantaneous
but fast response, essentially no phase
shift is induced on two-single-photon input states, and thus a quantum gate
built from such a medium cannot work. Here we show that a fast
response Kerr medium induces some but very little phase shifts on a two-single-photon input state, and it is insufficient for high fidelity quantum computation. We point out that this is caused by the medium imparting spectral entanglement to the input photons. We further show that
a way to circumvent this problem and achieve a high fidelity gate, is to engineer the dispersion
properties of the medium to give a dominant spectral effect over the
non-instantaneous response, in addition to satisfying a phase
matching condition.

\end{abstract}

\maketitle

\section{I. Introduction}

Research in quantum information sciences has shown that a large
scale quantum computer has the ability to factorize large
numbers, search through unsorted databases and perform quantum simulations, more efficiently than
classical computers. In order to achieve a universal set of quantum logic operations, quantum
entangling gates between qubits are required. Single photons have been demonstrated to be good carriers of quantum information. However, photons
interacts with each other weakly and this provides a major obstacle
in building an optical quantum computer. Nevertheless, various
schemes have been proposed to construct a two-qubit quantum
entangling gate. Linear optics with measurement induced
non-linearity can be used to build an entangling gate but it suffers
from a low success rate, which in other words means lots of resource are required to
achieve high efficiency~\cite{Knill01, Ralph06, Kok07}. This has led to renewed interests in
building a quantum gate with non-linear optics, which can provide
stronger interaction between photons~\cite{Milburn89}~\cite{Chuang95}. Recent developments in
non-linear optical quantum gates include gates such as the quantum Zeno gate~\cite{Franson04}~\cite{Leung06} and the Kerr
non-linearity gate~\cite{Nemoto04, Munro05}. 

\subsection{Cross-Kerr Quantum Gate}
\begin{figure}[h]
\centerline{\psfig{figure=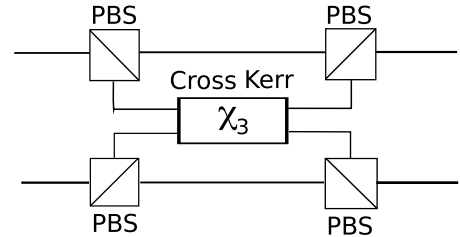,width=7cm}}
\caption{Schematic of the cross-Kerr controlled-sign gate in dual-rail logic. PBS stands for polarization beam splitters.\\} 
\label{fig:CZ}
\end{figure}

The Hamiltonian for a single frequency cross-Kerr medium with modes $p$ and $s$ can be written as:
\begin{equation}
\hat{H}=\chi\hat{a}_{p}^{\dagger}\hat{a}_{p}\hat{a}_{s}^{\dagger}\hat{a}_{s}
\end{equation}

\noindent where $\chi$ denotes the interaction strength. With a sufficiently strong cross-Kerr interaction, the Kerr medium can induce a $\pi$ phase shift on the two-single-photon state $|1\ra_p|1\ra_s$ without altering the $|0\ra_p|0\ra_s$ vacuum state, and the $|0\ra_p|1\ra_s$ and $|1\ra_p|0\ra_s$ single-photon states. For $|0\ra_j$ being logical zero and $|1\ra_j$ being  logical one, this is equivalent to a single rail logic controlled-sign gate. We note that since each mode may have at most one photon in our calculation, we do not include a self-phase modulation term in the Hamiltonian, because it will have zero contribution. For this Hamiltonian, the unitary operator is simply $\hat{U}=\exp\left(\frac{\hat{H}t}{i\hbar}\right)$. A $\pi$ phase shift on the $|11\rangle$ state can be obtained if $\frac{\chi t}{\hbar}=\pi$, and this gives the aforementioned controlled-sign operation. If we encode our qubits in polarization encoding, say $|H\ra$ is logical 0 and $|V\ra$ is logical 1, then this operation can be implemented as a gate in dual-rail logic~\cite{Lund02}, as shown in figure~\ref{fig:CZ}. Along with arbitrary single-qubit operations that can be done with waveplates~\cite{Dodd03}, universal quantum computation can be achieved.

Since the response of a non-linear medium is
frequency dependent, it is therefore important to know how its
spectral response affects the spectral profile of photonic input
states, as well as the fidelity of the gate. Recently,
Shapiro~\cite{Shapiro06} has argued that for a Kerr medium with
non-instantaneous but fast response, there is essentially no phase shift induced on the two-single-photon input states. Thus quantum gates based on such media cannot work. Furthermore, for a Kerr medium with slow response, significant phase shift can be imparted to the state but the presence of phase noise in this regime prevents a high fidelity operation. Here we show that a fast response Kerr medium does
induce some phase shifts on two-single-photon input states but insufficient for
high fidelity quantum computation. We identify the problem as predominately arising from spectral entanglement, that is, we show that for separable
two-single-photon input states, the output state from such a medium is
inevitably spectrally entangled. We then further show that a way to circumvent
the problem is to engineer the dispersion properties of the medium
to give a dominant spectral effect over the non-instantaneous
response, in addition to satisfying a phase matching condition.\\

This paper is arranged in the following way. The next section discusses how we model the spectral
properties of a $\chi^{(3)}$ cross-Kerr non-linear medium that has
a fast but non-instantaneous response. We then derive the output
state for a separable two photon input state and find the fidelity of the gate.
In Section III, we indicate how the
problem of spectral entanglement can be eliminated by introducing dispersion 
into the model and deduce a phase
matching condition that can enhance the fidelity. We conclude in section IV.

\section{II. Modelling the Spectral Response of a Kerr Medium with fast response}
The process of Kerr non-linearity can be studied in the interaction picture, with the
unitary evolution of a state vector given by:

\begin{equation}
\hat{U}(t_1,t_0)|\psi\rangle=\exp\left(\mathcal{T}\Big{\{}-\frac{i}{\hbar}\int^{t_1}_{t_0}\hat{H}(t)dt\Big{\}}\right)|\psi\rangle
\label{eq:U}
\end{equation}

\noindent where $\mathcal{T}$ is the time ordering operator. Since the interaction Hamiltonian for a cross-Kerr non-linearity does not necessarily commute at different times, therefore the Dyson series~\cite{Dyson, Sakurai} shall be used to calculate the higher order terms of the unitary expansion. In the case of a time-commutable interaction Hamiltonian, the time ordering operator has no effect and may be dropped, resulting in the usual Taylor series for the unitary expansion. The interaction Hamiltonian for a cross-Kerr non-linear medium with non-instantaneous response and no self Kerr effect, has the expression as shown in equation~(\ref{eq:rawH}).

\begin{align}
\hat{H}(t)=&\chi^{(3)}\epsilon_{0}\int_{V}d\mathbf{r}^{3}\Big{(}\hat{E_p}^{\dagger}(\mathbf{r},t)\hat{E_p}(\mathbf{r},t)\Big{(}\int^{\infty}_{0}d\tau g(\tau)\nn\\
&\hat{E_s}^{\dagger}(\mathbf{r},t-\tau)\hat{E_s}(\mathbf{r},t-\tau)+\kappa\hat{m}_p(\tau,r)\Big{)}\nn\\
	+&\Big{(}\int^{\infty}_{0}d\tau g(\tau')\hat{E_p}^{\dagger}(\mathbf{r},t-\tau')\hat{E_p}(\mathbf{r},t-\tau')\nn\\
	&+\kappa\hat{m}_s(\tau',r)\Big{)}\hat{E_s}^{\dagger}(\mathbf{r},t)\hat{E_s}(\mathbf{r},t)\Big{)}
\label{eq:rawH}
\end{align}

Hereafter, we shall limit the analysis to one spatial dimension, the propagation direction. The expression for the electric field operator of mode $j$ is:

\begin{equation}
\hat{E}^{\dagger}_{j}(z,t)=\int^{\infty}_{-\infty}d\omega_{j}A_j(\omega_j)\hat{a_j}^{\dagger}(\omega_j)e^{i (k_j(\omega_j)z-\omega_j t)}
\end{equation}

\noindent where $A_j(\omega_j)=i\sqrt{\frac{\hbar\omega_j}{4\pi c\epsilon_0
n^2_j(\omega_0)S}}$ and $S$ is the cross section area of the beam and $n_j(\omega_j)$ is the refractive index for mode $j$.
We assume that $A_j(\omega_j)=A_j$ is slowly varying for the
frequencies of interest, allowing it to be factored outside the
integral. In this section, we are interested in the effects of non-instantaneous response, and so we assume that it dominates over the effects of dispersion and hence we assume wavenumber $k_j$ is constant.\\

The $\hat{m}(\tau,z)$ terms in the Hamiltonian are noise operators that are required to preserve the commutation of the field operators when the medium has a non-instantaneous response~\cite{Blow91, Boivin94, Shapiro06}. In the case of self-phase modulation, Boivin et al~\cite{Boivin94} showed that the noise term can be modelled as a collection of localized and independent harmonic oscillators with the expression:

\begin{equation}
\hat{m}(\tau,z)=\int_{0}^{\infty}d\omega\sqrt{\frac{G(\omega)}{\pi}}\hat{d}^{\dagger}_{\omega}(z)e^{i\omega\tau}+H.C.
\label{eq:noise}
\end{equation}

\noindent where $G(\omega)=\int d\tau \textrm{sin}(\omega\tau)g(\tau)$. Furthermore, in the case of under-damping, i.e. $0<\Gamma<2\Omega$, the response function for $\tau>0$ is 

\begin{equation}
g(\tau)=\frac{K\Omega^2}{\sqrt{\Omega^2-\Gamma^{2}/4}}e^{-\Gamma\tau/2}\textrm{sin}\left(\sqrt{\Omega^2-\Gamma^{2}/4}\tau\right)
\end{equation}

\noindent where $\Gamma$ is the damping coefficient and $\Omega$ is the resonance frequency of the medium. Shapiro~\cite{Shapiro06} commented that this model of noise also applies to cross-phase modulation. If the medium has a very high resonance frequency and a large damping coefficient, $g(\tau)$ resembles a delta function, which gives the fast response limit that we are interested in. To illustrate this, for instance, lets consider the limit of critical damping, $\Gamma=2\Omega$, and such that $g(\tau)=K\Omega^2\tau e^{-\Omega\tau}$. The maximum of this function is $K\Omega e^{-1}$ and occurs at $\tau=1/\Omega$. Furthermore, the full width half maximum is proportional to $1/\Omega$. Hence, for large $\Omega$, the response function resembles a delta function. For mathematical convenience in our calculation, in the fast response regime, we approximate $g(\tau)\approx\frac{M}{\sqrt{\pi}}\exp(-M^2\tau^2)$ as a narrow gaussian function in time with $\tau>0$ and large $M$. Here, $M\propto\Omega$ and that in the limit where $M\rightarrow\infty$, we have $g(\tau)\rightarrow\delta(\tau)$. As the response function tends to a delta function, equation~(\ref{eq:noise}) shows that the noise term tends to zero and therefore in the fast response regime, the noise term has negligible contribution and the noise term in the Hamiltonian shown in equation~(\ref{eq:rawH}) can be ignored.\\

For the Hamiltonian, after integrating over $\tau$ by assuming large M, and similarly for the $s$ mode, in the expression we have the term:
\begin{equation}
\int^{\infty}_{0}d\tau g(\tau)e^{-i(\omega_{p+}-\omega_{p-})\tau}\approx\frac{1}{2}e^{-\frac{(\omega_{p+}-\omega_{p-})^2}{4M^2}}
\end{equation}

Thus the interaction Hamiltonian has the expression:  
\begin{align}
	\hat{H}(t)\approx&\frac{\chi L}{2}\iiiint_{-\infty}^{\infty} d\omega_{p+}d\omega_{p-}d\omega_{s+}d\omega_{s-}\nn\\
	&\hat{a}_{p}^{\dagger}(\omega_{p+})\hat{a}_{p}(\omega_{p-})\hat{a}_{s}^{\dagger}(\omega_{s+})\hat{a}_{s}(\omega_{s-})\exp(-i\Delta\omega t))\nn\\
	&\left(e^{-\frac{(\omega_{p+}-\omega_{p-})^{2}}{4M^{2}}} + e^{-\frac{(\omega_{s+}-\omega_{s-})^{2}}{4M^{2}}}\right)
\label{eq:H1}
\end{align}

\noindent where $\Delta\omega=\omega_{p+}-\omega_{p-}+\omega_{s+}-\omega_{s-}$ is the frequency detuning. $L$ is the length of the medium, which comes from integrating $z$ from 0 to $L$. The integration of $z$ does not involve the phases associated with the wavenumber of the modes because the sum of the wavenumbers is zero when dispersion is assumed to be negligible. Here $\chi$ is again the interaction strength but incorporated with some constants from the electric field expressions. The frequency integrals have lower bounds extended from zero to negative infinity. This is mathematically legitimate because we are considering a system that operates at high frequency, where essentially there is no population present at low frequency.\\

This interaction Hamiltonian does not commute at different times, so strictly speaking, we should use the Dyson series to calculate higher order terms in the unitary expansion. However, here we shall use the Taylor series as an approximation to the Dyson series. Later in this section, we shall show with the Taylor series that the non-instantaneous response of the Kerr medium will always produce spectral entanglement and poses a problem to the fidelity the gate. Because of this, for a non-instantaneous Kerr medium, we expect the more complicated Dyson series will also show a similar spectral entanglement problem. In the next section, we shall see that engineering the dispersion property of the medium can help eliminate the spectral entanglement and enhance the gate fidelity. Hence, for the purpose of illustrating the spectral entanglement problem, it is sufficient to use the Taylor series, even though formally we should use the Dyson series. Nonetheless, in the appendix, we will further discuss the validity of this approximation.\\ 

Integrating the Hamiltonian in equation~(\ref{eq:H1}) over time $t$ with bounds taken from $-\infty$ to $\infty$ by considering far field limits~\cite{Giovannetti02}, results in a factor of $\delta(\Delta\omega)$. When we apply the energy conservation condition by integrating $\delta(\Delta\omega)$, the Hamiltonian becomes~\footnote{In the case of energy conservation, the Hamiltonian has the form of having a frequency filtering function for only the $p$ mode but not the $s$ mode, thus there is no functional difference between having a time delay in one mode or both}:

\begin{align}
	\hat{H}=&\int_{-\infty}^{\infty}\hat{H}(t)dt\nn\\
	=&\chi L\iiint d\omega_{p+}d\omega_{p-}d\omega_{s+}\hat{a}_{p}^{\dagger}(\omega_{p+})\hat{a}_{p}(\omega_{p-})\hat{a}_{s}^{\dagger}(\omega_{s+})\nn\\
	&\hat{a}_{s}(\omega_{p+}-\omega_{p-}+\omega_{s+})\exp\left(-\frac{(\omega_{p+}-\omega_{p-})^{2}}{4M^{2}}\right)
\label{eq:H2}
\end{align}

\subsection{Effects of non-instantaneous response}
We have derived an expression for a Kerr medium with finite response time as shown in equation~(\ref{eq:H2}). The finite response time translates into a spectral filtering function in frequency space. Here we shall see what happens when we apply this cross-Kerr process to two-single-photon input states. A single-photon state with a gaussian spectral profile is: 
\begin{equation}
	|1\rangle=\sqrt{\frac{1}{\sigma\sqrt{2\pi}}}\int_{-\infty}^{\infty} d\omega~\hat{a}^{\dagger}(\omega)\exp\left(-\frac{\nu^{2}}{4\sigma^{2}}\right)|0\rangle
\end{equation}

So the spectrally separable input state with modes $p$ and $s$ is: 
\begin{align}
	|11\rangle=&\frac{1}{\sigma\sqrt{2\pi}}\iint_{-\infty}^{\infty} d\omega_{p}d\omega_{s}\hat{a}_{p}^{\dagger}(\omega_{p})\hat{a}_{s}^{\dagger}(\omega_{s})\nn\\
	&\exp \Big{(}-\frac{\nu_{p}^{2}}{4\sigma^{2}} \Big{)} \exp \Big{(} -\frac{\nu_{s}^{2}}{4\sigma^{2}} \Big{)} |0\rangle
\end{align}

\noindent where $\nu_{j}=\omega_{j}-\mu$ and $\mu$ is the mean frequency. Having the unitary acting on $|11\rangle$ gives:
\begin{equation}
\hat{U}|11\rangle=\sum_{n=0}^{\infty}\frac{1}{n!}\left(\frac{\hat{H}}{i\hbar}\right)^{n}|\psi_{0}\rangle=\sum_{n=0}^{\infty}|\psi_{n}\rangle
\label{eq:psi}
\end{equation}

Note that $|\psi_0\ra=|11\ra$. From equation~(\ref{eq:H2}) and ~(\ref{eq:psi}), one can show that 
\begin{align}
|\psi_{n}\rangle=&\frac{1}{n!}\left(\frac{\chi L}{i\hbar}\right)^{n}2^{\frac{2n-1}{2}}\pi{}^{\frac{n-1}{2}}\sqrt{\frac{M^{2n}}{2nM^{2}+\sigma^{2}}}\nn\\
&\iint d\omega_{p}d\omega_{s}\hat{a}_{p}^{\dagger}(\omega_{p})\hat{a}_{s}^{\dagger}(\omega_{s})\nn\\
&\exp\left(-\frac{nM^{2}(\nu_{p}+\nu_{s})^{2}+\sigma^{2}(\nu_{p}^{2}+\nu_{s}^{2})}{4\sigma^{2}(2nM^{2}+\sigma^{2})}\right)|0\rangle
\label{eq:psi_n}
\end{align}

Expression~(\ref{eq:psi_n}) implies that cross-Kerr non-linearity inevitably induces spectral entanglement in the output state because the cross term in the argument of the exponential $\frac{2nM^{2}\nu_{p}\nu_{s}}{4\sigma^{2}(2nM^{2}+\sigma^{2})}$ cannot vanish for non-zero $M$. This is undesirable for quantum computation, as entanglement in the spectral domain will appear as decoherence in the logical basis, thus lowering the fidelity.\\ 

Given the input state $|\psi_{in}\rangle=(|10\rangle+|11\rangle)/\sqrt{2}$, the expected state is 
$|\psi_{expected}\rangle=(|10\rangle-|11\rangle)/\sqrt{2}$. The actual output state is:
\begin{equation}
|\psi_{out}\rangle=\hat{U}|\psi_{in}\rangle=\frac{1}{\sqrt{2}}\left(|10\rangle+\sum_{n=0}^{\infty}|\psi_{n}\rangle\right)
\end{equation}

\begin{figure}[h]
\centerline{\psfig{figure=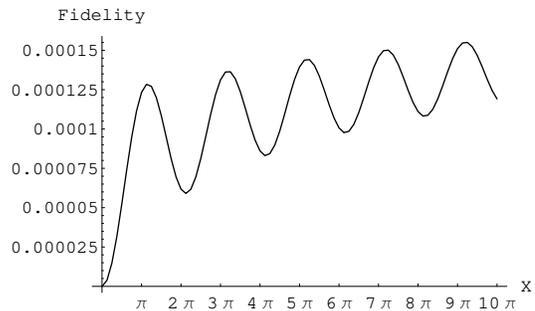,width=7cm}}
\caption{Fidelity vs $X$ for $\eta=0.01$.\\}
\label{fig:FvsX}
\end{figure}

\begin{figure}[h]
\centerline{\psfig{figure=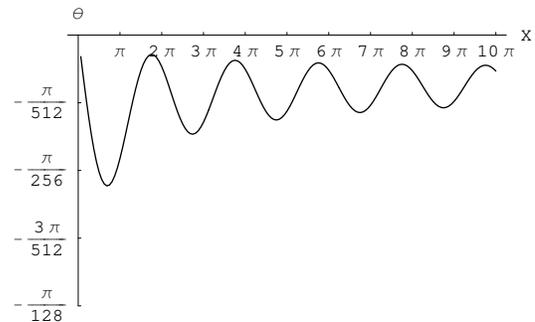,width=7cm}}
\caption{$\theta$ vs $X$ for $\eta=0.01$. Note that the argument is undetermined at $X$=0 because the complex number $\langle\psi_{expected}|\psi_{out}\rangle=0$.\\}
\label{fig:ThetavsX}
\end{figure}

Hence the expression for the fidelity is:
\begin{align}
F=&\frac{1}{4}\left|1-\sum_{n=0}^{\infty}\langle 11|\psi_{n}\rangle\right|^{2}\nn\\
=&\frac{1}{4}\left|1-\sum_{n=0}^{\infty}\frac{1}{n!}\left(\frac{2\sqrt{\pi}\chi L}{i\hbar}\right)^{n}\sqrt{\frac{M^{2n}\sigma^{2}}{nM^{2}+\sigma^{2}}}\right|^{2}\nn\\
=&\frac{1}{4}\left|1-\sum_{n=0}^{\infty}\frac{\left(-iX\right)^{n}}{n!}\frac{\eta}{\sqrt{n+\eta^2}}\right|^{2}
\label{eq:fidelity}
\end{align}

\noindent where $\eta=\sigma/M$ and $X=\frac{2\sqrt{\pi}\chi L M}{\hbar}$. In the case of fast response, i.e. $\eta \ll 1$, equation~(\ref{eq:fidelity}) becomes $F=\frac{1}{4}\Big{|}\sum_{n=1}^{\infty}\frac{1}{n!}\left(-iX\right)^{n}\frac{\eta}{\sqrt{n}}\Big{|}^{2}$~\footnote{The numerical sum in the expression for the fidelity goes up to $n=200$, as higher order terms have negligible contribution}. To see the effect of the frequency response function on the state, we shall find the fidelity as a function of $X$ and $\eta$. In particular, we set $\eta=0.01$ for Figure~\ref{fig:FvsX}. Notice in the figure that even when $X$ is large, the maximum fidelity is roughly in the order of $10^{-4}$. Similarly, for $\eta=0.001$, the maximum fidelity is on the order of $10^{-6}$ for large $X$. Hence the fidelity is far too low for quantum computation in the fast response regime. The low fidelity implies that the output state is almost orthogonal to the expected state. It also implies that the fidelity of the output state with respect to the input state is near unity. Figure~\ref{fig:ThetavsX} plots $\theta$ against $X$, where $\theta$ is the argument of the complex number $\la11|\hat{U}|11\ra$ in polar form~\footnote{The numerical sum only goes up to $n=200$, as higher order terms are negligible}. If the cross-Kerr medium does induce a $\pi$ phase shift onto the two-single-photon state, then $\theta$ would be $\pm\pi$. However, the figure shows that for large $X$, $\theta$ is roughly $-\frac{\pi}{512}$, which indicates some but very little phase is induced onto the state by the medium.\\

To understand the fact that it is the spectral entanglement that lowers the fidelity, we shall examine the case where the spectral entanglement is negligible. Which is the case of slow response, where $M$ is small but conveniently ignoring the noise for argument sake. Let us suppose that the Hamiltonian has the same form as equation~(\ref{eq:H2}), where the noise term is ignored, and also assume that the higher order terms in the summation of equation~(\ref{eq:fidelity}) is only significant up to the $n_{c}$-th term. If we consider the regime where $\eta \gg n_{c}>1$, then we can see from equation~(\ref{eq:psi_n}) that the cross term becomes insignificant, and $F=\frac{1}{4}\left|1-\sum_{n=0}^{n_{c}}\frac{1}{n!}\left(-iX\right)^{n}\right|^{2}$. Setting $X=\pi$ gives $F = 1$, which means a $\pi$ phase shift is induced on $|11\rangle$. Hence, we argue that it is the spectral entanglement caused by the non-instantaneous response function in the fast regime that leads to a low fidelity. Although this entanglement is reduced in the slow response regime, the presence of noise terms there reduce the fidelity of operation there as well, as pointed out by Shapiro~\cite{Shapiro06}. We seek a way to reduce the spectral entanglement whilst remaining in the fast response regime.

\section{III The spectral response of a Kerr medium with dispersion}

In the previous section, we have discussed what happens if we input a separable two-photon state into a Kerr medium with finite response time. In the fast response regime, the spectral entanglement induced by the non-instantaneous response function reduces the phase shift imparted and thus lowers the fidelity. Hence if we can engineer the phase such that the spectral entanglement can be eliminated, then a unit fidelity may again be achieved. It is well known that by adjusting the dispersion in a parametric down conversion crystal, the spectral profile of the output photons can be modified~\cite{Grice97}. Here we propose to engineer the dispersion properties of a Kerr medium such that the resultant state is a separable two-photon state. Furthermore, with a special condition set for $\chi$, we show that the resultant state can have a $\pi$ phase shift, which is what we hope for.\\

The interaction Hamiltonian with dispersion and non-instantaneous response is:
\begin{align}
	\hat{H}(t)\approx&\chi L\iiiint_{-\infty}^{\infty} d\omega_{p+}d\omega_{p-}d\omega_{s+}d\omega_{s-}\nn\\
	&\hat{a}_{p}^{\dagger}(\omega_{p+})\hat{a}_{p}(\omega_{p-})\hat{a}_{s}^{\dagger}(\omega_{s+})\hat{a}_{s}(\omega_{s-})\nn\\
	&\left(\exp(-\frac{(\omega_{p+}-\omega_{p-})^{2}}{4M^{2}}) + \exp(-\frac{(\omega_{s+}-\omega_{s-})^{2}}{4M^{2}})\right)\nn\\
& \textrm{sinc}\left(\frac{L\Delta k}{2}\right)\exp\left(\frac{iL\Delta k}{2}\right)\exp(-i\Delta\omega t)
\label{eq:Hd}
\end{align}

\noindent where $\Delta k\approx k_{p}^{'}(\omega_{p+}-\omega_{p-})+k_{s}^{'}(\omega_{s+}-\omega_{s-})$ is the difference in the wavenumber of mode $p$ and $s$, i.e. the phase mismatch, up to the first order terms after a Taylor series expansion about the mean frequency. Here $k'_{j}$ is the derivative of the wavenumber of mode $j$ with respect to $\omega_j$ and evaluated at the mean frequency. We have assumed that the higher order terms of $\Delta k$ are negligible. Moreover, the zero-th order term is zero because of conservation of momentum. Later in the appendix, we shall show that when the spectral effects of the dispersion dominates over the non-instantaneous effect, the interaction Hamiltonian does commute at different times, and thus we may use the Taylor series to calculate higher order terms of the unitary expansion. Hence the n-th state in the sum now has the expression:

\begin{align}
|\psi_{n}\rangle=&\sqrt{\frac{M^{2n}\mathcal{L}^{2n}(\sigma^{2}+M^{2}\mathcal{L}^{2})}{(\sigma^{2}+M^{2}\mathcal{L}^{2})^{n}(2nM^{2}+\sigma^{2}+M^{2}\mathcal{L}^{2})}}\nn\\
&\frac{1}{n!}\left(\frac{\pi}{i}\right)^{n}\frac{1}{\sigma\sqrt{2\pi}}\iint d\omega_{p}d\omega_{s}\hat{a}_{p}^{\dagger}(\omega_{p})\hat{a}_{s}^{\dagger}(\omega_{s})\nn\\
&e^{-(C_{1}\nu_{p}^{2}+C_{1}\nu_{s}^{2}+C_{2}\nu_{p}+C_{2}\nu_{s}+C_{3}\nu_{p}\nu_{s}+C_{4})}|0\rangle
\label{eq:psi_n_dispersion}
\end{align}

\noindent where we define $\mathcal{L}=L\sigma\sqrt{\gamma(k_{p}^{'}-k_{s}^{'})^{2}}$,\\
$C_{1}={\frac{nM^{2}+\sigma^{2}+M^{2}\mathcal{L}^{2}}{4\sigma^{2}(2nM^{2}+\sigma^{2}+M^{2}\mathcal{L}^{2})}}$,
$C_{2}=\frac{in\mathcal{L}M^{2}}{2\sigma\sqrt{\gamma}(2nM^{2}+\sigma^{2}+M^{2}\mathcal{L}^{2})}$
$C_{3}=\frac{nM^{2}}{2\sigma^{2}(2nM^{2}+\sigma^{2}+M^{2}\mathcal{L}^{2})}$, and
$C_{4}=\frac{n(2nM^{2}+\sigma^{2})}{4\gamma(2nM^{2}+\sigma^{2}+M^{2}\mathcal{L}^{2})}$\\

Here we have approximated the sinc function of the phase matching function as a gaussian function via the approximation $sinc(x)\approx\exp(-\gamma x^2)$ with $\gamma\approx0.193...$ calculated from matching the full-width-half-maximum of the two functions. From equation~(\ref{eq:psi_n_dispersion}), we can see that in the limit of $\mathcal{L}\rightarrow\infty$, the cross term $C_{3}$ is zero and thus the resultant state is separable. Along with a special condition on $\chi$ that allows the unitary to give a $\pi$ phase shift, we could achieve unit fidelity. It is therefore of interest to know what length of material can give a high fidelity. The new expression for the fidelity is:

\begin{align}
F=&\sum_{n=0}^{\infty}-\frac{1}{n!}\left(-\frac{2\pi i\chi L}{\hbar}\right)^{n}\nn\\
&\sqrt{\frac{M^{2n}\gamma^{n}\sigma^{2}(1+L^{2}M^{2}\gamma(k_{p}^{'}-k_{s}^{'})^{2})^{1-n}}{M^{2}n+\sigma^{2}+L^{2}M^{2}\gamma\sigma^{2}(k_{p}^{'}-k_{s}^{'})^{2}}}\nn\\
&\exp\left(-\frac{nL^{2}M^{2}\sigma^{2}(k_{p}^{'}-k_{s}^{'})^{2}}{4(M^{2}n+\sigma^{2}+L^{2}M^{2}\gamma\sigma^{2}(k_{p}^{'}-k_{s}^{'})^{2})}\right)
\label{eq:Feq}
\end{align}

\begin{figure}[h]
\centerline{\psfig{figure=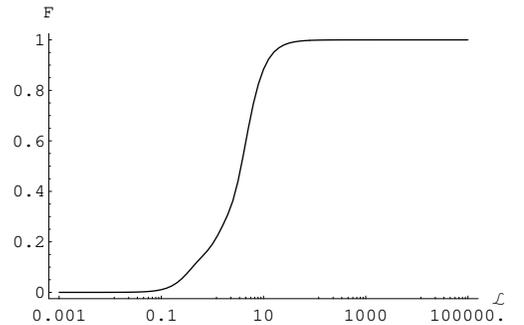,width=7cm}}
\caption{Fidelity vs $\mathcal{L}$.\\}
\label{fig:FvsL}
\end{figure}

We now set the special condition $\frac{2\chi}{\hbar|k_{p}^{'}-k_{s}^{'}|}=\exp(\frac{1}{4\gamma})$, which is derived from taking the limit $L\rightarrow\infty$, such that the unitary induces a $\pi$ phase shift on $|11\rangle$, in other words, $F=1$ in equation~(\ref{eq:Feq}). With this condition, we simplify the expression for the fidelity to:

\begin{equation}
F=\frac{1}{4}\Big{|}1-\sum_{n=0}^{\infty}\frac{(-i\pi)^{n}\mathcal{L}}{n!\sqrt{n+\mathcal{L}^{2}}}\exp\left(\frac{n^{2}}{4\gamma(n+\mathcal{L^{2}})}\right)\Big{|}^{2}
\end{equation}

\noindent where we have assumed $\frac{1}{M^{2}}\ll L^{2}\gamma(k_{p}^{'}-k_{s}^{'})^{2}$, i.e. the effect of dispersion dominates over the effect of non-instantaneous response. Figure~\ref{fig:FvsL} is the plot of fidelity versus the dimensionless $\mathcal{L}$. When $F=0.999$, we have $\mathcal{L}=100$, so for the reasonable parameters $\Big{|}k_{p}^{'}-k_{s}^{'}\Big{|}=10^{-8} (s/m)$ and $\sigma=10^{13} (Hz)$, the corresponding length of the medium is $L=2.28*10^{-3} (m)$.

\section{IV. Conclusion}
In this paper, we have modeled the spectral effects of a non-instantaneous cross-Kerr medium and discussed how the noise terms of the interaction Hamiltonian can be ignored in the fast response regime. We have shown that for a separable input of two-single-photon state, a fast response cross-Kerr medium inevitably spectrally entangles the state at output. Ideally, a cross-Kerr medium would impose a $\pi$ phase shift on the input state, but we have shown that the presence of the entanglement reduces the phase shift to very little, and found that the resulting fidelity is too low for quantum computation. We have illustrated that a way to circumvent
the problem is to engineer the dispersion properties of a fast response medium
to give a dominant spectral effect, in addition to satisfying the phase matching condition $L\sigma\sqrt{\gamma(k_{p}^{'}-k_{s}^{'})^{2}}\rightarrow\infty$. The dispersion helps the state to remain separable, and by further picking a particular value of $\chi$, a $\pi$ phase shift can be induced on the two-single-photon state.  We therefore argue that it is possible to build a high
fidelity cross-Kerr quantum gate from a single pass medium of sufficient non-linearity, provided one could engineer the dispersion of a fast response medium. Our calculation shows that in order to achieve a fidelity of 0.999, we need the dimensionless parameter $\mathcal{L}=100$. An example is where the difference of the reciprocal of the group velocities is $10^{-8}(s/m)$ and the spectral width of the photons is $10^{13} (Hz)$, and a length of medium of $2.28(mm)$.\\

\textbf{Acknowledgement}\\

\indent We thank Andrew White for useful discussions.
This work was supported by the Australian Research Council and the
IARPA-funded U.S. Army Research Office Contract No. W911NF-05-0397. We would also like to acknowledge support from the EU Integrated Project Qubit Applications (QAP) and the Ministry of Education, Culture, Sports, Science and Technology (MEXT) of Japan. P. Leung would also like to give special thanks to NII for their hospitality while a portion of this research was conducted there.

\section{Appendix}

\subsection{A1: Approximating the Dyson Series with Taylor Series for the Interaction Hamiltonian with Non-Instantaneous Response}

The interaction Hamiltonian with a non-instantaneous response does not commute at different times and strictly speaking, we should use the Dyson series instead of the Taylor series to calculate higher order terms of the unitary expansion. However, we have calculated the fidelity for second order term of the Dyson and Taylor series with various values of $M$ and $\sigma$ of interest and the fidelity seems to be close to unity. In particular, when $M=10^{10}$ and $\sigma=10^9$, the fidelity is approximately 1.046 with a roughly estimated numerical error of $\pm0.077$. Fidelity is always bounded by one and this bound is within the numerical error estimated. Hence arguably, the second order term of the Taylor and Dyson series are identical. This suggests that the Taylor series may possibly give a good approximation for higher order terms of the evolution of spectrally-Gaussian input states passing through a bulk cross Kerr medium.\\

Furthermore, if the Kerr medium consists of thin slices separated sufficiently apart, such that the wavepacket exits one slice before entering another, and that each slice only provides a weak interaction, then it is appropriate to apply the Taylor series instead of the Dyson series for the unitary operation expansion. This is because the unitary operator can be expressed as equation~(\ref{eq:U2}) and if the slices are sufficiently apart, the time bounds of the  integrals can be conveniently extended to $-\infty$ and $\infty$ by considering far field limits~\cite{Giovannetti02}, and thus each factor in the series is the same, and the time ordering operator can again be dropped, which gives the Taylor series expansion.

\begin{align}
\hat{U}(t_1,t_0)=&\lim_{n\rightarrow\infty}\mathcal{T}\Big{\{}\left(1-\frac{i}{\hbar}\int^{t_1}_{t_1-\Delta T}\hat{H}(t)dt\right)\nn\\
&\left(1-\frac{i}{\hbar}\int^{t_1-\Delta T}_{t_1-2\Delta T}\hat{H}(t)dt\right)\dots\nn\\
&\left(1-\frac{i}{\hbar}\int^{t_0+\Delta T}_{t_0}\hat{H}(t)dt\right)\Big{\}}
\label{eq:U2}
\end{align}

\noindent where $\Delta T=(t_1-t_0)/n$.\\

\subsection{A2: Time Commutivity of the Interaction Hamiltonian with Dispersion Effects}
Here we shall prove that in the fast response limit with a dominant effect from the dispersion of the medium, the Hamiltonian in equation~(\ref{eq:Hd}) commutes at different times and the expression for the unitary operator is properly given in equation~(\ref{eq:U}) without the time ordering operator.\\

In the limit of very large $M$, the interaction Hamiltonian in equation~(\ref{eq:Hd}) tends to:
\begin{align}
	\hat{H}(t)\rightarrow&\chi L\iiiint d\omega_{p+}d\omega_{p-}d\omega_{s+}d\omega_{s-}\hat{a}_{p}^{\dagger}(\omega_{p+})\hat{a}_{p}(\omega_{p-})\nn\\
	&\hat{a}_{s}^{\dagger}(\omega_{s+})\hat{a}_{s}(\omega_{s-})\textrm{sinc}\Big{(}\frac{L\Delta k}{2}\Big{)}e^{i\frac{L\Delta k}{2}}e^{-i\Delta\omega t}
\end{align}

Hence the commutation of the interaction Hamiltonian at times $t_1$ and $t_2$ is:

\begin{align}
	&[\hat{H}(t_1),\hat{H}(t_2)]\nn\\
	=&\chi^2 L^2\int\dots\int d\omega_{p+}d\omega_{p-}d\omega_{s+}d\omega_{s-}d\omega_{p+}'d\omega_{p-}'d\omega_{s+}'d\omega_{s-}'&\nn\\
	&\Big{[}\hat{a}_{p}^{\dagger}(\omega_{p+})\hat{a}_{p}(\omega_{p-})\hat{a}_{s}^{\dagger}(\omega_{s+})\hat{a}_{s}(\omega_{s-}),\nn\\
	&\hat{a}_{p}^{\dagger}(\omega_{p+}')\hat{a}_{p}(\omega_{p-}')\hat{a}_{s}^{\dagger}(\omega_{s+}')\hat{a}_{s}(\omega_{s-}')\Big{]}e^{-i\Delta\omega t_1}e^{-i\Delta\omega' t_2}\nn\\
	&\textrm{sinc}\Big{(}\frac{L\Delta k}{2}\Big{)}\textrm{sinc}\Big{(}\frac{L\Delta k'}{2}\Big{)}e^{i\frac{L\Delta k}{2}}e^{i\frac{L\Delta k'}{2}}
	\label{eq:HCommutation}
\end{align}

\noindent where $\Delta k'\approx k_{p}'(\omega_{p+}'-\omega_{p-}')+k_{s}'(\omega_{s+}'-\omega_{s-}')$ and\\ $\Delta\omega'=\omega_{p+}'-\omega_{p-}'+\omega_{s+}'-\omega_{s-}'$.\\

The commutation relation in the above equation can be expressed as a sum of 8 terms:

\begin{align}
&\Big{[}\hat{a}_{p}^{\dagger}(\omega_{p+})\hat{a}_{p}(\omega_{p-})\hat{a}_{s}^{\dagger}(\omega_{s+})\hat{a}_{s}(\omega_{s-}),\nn\\
	&\hat{a}_{p}^{\dagger}(\omega_{p+}')\hat{a}_{p}(\omega_{p-}')\hat{a}_{s}^{\dagger}(\omega_{s+}')\hat{a}_{s}(\omega_{s-}')\Big{]}\nn\\
	=&-\Big{(}\delta{}_{s+,s'-}\hat{a}_{s}^{\dagger}(\omega_{s+}')\hat{a}_{s}(\omega_{s-})\hat{a}_{p}^{\dagger}(\omega_{p+})\hat{a}_{p}(\omega_{p-})\hat{a}_{p}^{\dagger}(\omega'_{p+})\hat{a}_{p}(\omega_{p-}')\nn\\
	&-\delta{}_{s'+,s-}\hat{a}_{s}^{\dagger}(\omega_{s+})\hat{a}_{s}(\omega_{s-}')\hat{a}_{p}^{\dagger}(\omega_{p+})\hat{a}_{p}(\omega_{p-})\hat{a}_{p}^{\dagger}(\omega_{p+}')\hat{a}_{p}(\omega_{p-}')\nn\\
	&\delta{}_{p+,p'-}\hat{a}_{p}^{\dagger}(\omega_{p+}')\hat{a}_{p}(\omega_{p-})\hat{a}_{s}^{\dagger}(\omega_{s+})\hat{a}_{s}(\omega_{s-})\hat{a}_{s}^{\dagger}(\omega_{s+}')\hat{a}_{s}(\omega_{s-}')\nn\\
	&-\delta{}_{p'+,p-}\hat{a}_{p}^{\dagger}(\omega_{p+})\hat{a}_{p}(\omega'_{p-})\hat{a}_{s}^{\dagger}(\omega_{s+})\hat{a}_{s}(\omega_{s-})\hat{a}_{s}^{\dagger}(\omega_{s+}')\hat{a}_{s}(\omega_{s-}')\nn\\
	&\delta{}_{p'+,p-}\hat{a}_{p}^{\dagger}(\omega_{p+})\hat{a}_{p}(\omega'_{p-})\delta{}_{s'+,s-}\hat{a}_{s}^{\dagger}(\omega_{s+})\hat{a}_{s}(\omega'_{s-})\nn\\
	&-\delta{}_{p'+,p-}\hat{a}_{p}^{\dagger}(\omega_{p+})\hat{a}_{p}(\omega'_{p-})\delta{}_{s+,s'-}\hat{a}_{s}^{\dagger}(\omega'_{s+})\hat{a}_{s}(\omega_{s-})\nn\\
	&\delta{}_{p+,p'-}\hat{a}_{p}^{\dagger}(\omega'_{p+})\hat{a}_{p}(\omega{}_{p-})\delta{}_{s+,s'-}\hat{a}_{s}^{\dagger}(\omega'_{s+})\hat{a}_{s}(\omega{}_{s-})\nn\\
	&-\delta{}_{p+,p'-}\hat{a}_{p}^{\dagger}(\omega'_{p+})\hat{a}_{p}(\omega{}_{p-})\delta{}_{s'+,s-}\hat{a}_{s}^{\dagger}(\omega{}_{s+})\hat{a}_{s}(\omega'_{s-})\Big{)}
\end{align}

\noindent where $\delta_{x,y}=\delta(\omega_x-\omega_y)$. After substituting this into equation~(\ref{eq:HCommutation}), we will also obtain 8 terms. If we interchange the variables $\omega_{s+}'\leftrightarrow\omega_{s+}$ and $\omega_{s-}'\leftrightarrow\omega_{s-}$ for the first term and then add with the second term, we obtain:

\begin{align}
	&[\hat{H}(t_1),\hat{H}(t_2)]\nn\\
	=&-\chi^2 L^2\int\dots\int d\omega_{p+}d\omega_{p-}d\omega_{s+}d\omega_{s-}d\omega_{p+}'d\omega_{p-}'d\omega_{s+}'d\omega_{s-}'&\nn\\
	&\hat{a}_{s}^{\dagger}(\omega_{s+})\hat{a}_{s}(\omega_{s-}')\hat{a}_{p}^{\dagger}(\omega_{p+})\hat{a}_{p}(\omega_{p-})\hat{a}_{p}^{\dagger}(\omega_{p+}')\hat{a}_{p}(\omega_{p-}')\delta{}_{s'+,s-}\nn\\	
	&\Big{(}e^{-i(\omega_{p+}-\omega_{p-}+\omega_{s+}'-\omega_{s-}') t_1}e^{-i(\omega_{p+}'-\omega_{p-}'+\omega_{s+}-\omega_{s-}) t_2}\nn\\
		&\textrm{sinc}\Big{(}\frac{L(\Delta k'_{p'}+\Delta k'_{s})}{2}\Big{)}e^{\frac{i L}{2}(\Delta k'_{p'}+\Delta k'_{s})}\nn\\
		&\textrm{sinc}\Big{(}\frac{L(\Delta k'_p+\Delta k'_{s'})}{2}\Big{)}e^{\frac{i L}{2}(\Delta k'_p+\Delta k'_{s'})}\nn\\
	&-e^{-i(\omega_{p+}-\omega_{p-}+\omega_{s+}-\omega_{s-}) t_1}e^{-i(\omega_{p+}'-\omega_{p-}'+\omega_{s+}'-\omega_{s-}') t_2}\nn\\
		&\textrm{sinc}\Big{(}\frac{L(\Delta k'_{p'}+\Delta k'_{s'})}{2}\Big{)}e^{\frac{i L}{2}(\Delta k'_{p'}+\Delta k'_{s'})}\nn\\
		&\textrm{sinc}\Big{(}\frac{L(\Delta k'_p+\Delta k'_s)}{2}\Big{)}e^{\frac{i L}{2}(\Delta k'_p+\Delta k'_s)}\Big{)}+\dots
\end{align}

\noindent where $\Delta k'_p=k_{p}'(\omega_{p+}-\omega_{p-})$, $\Delta k'_{s}=k_{s}'(\omega_{s+}-\omega_{s-})$, $\Delta k'_{p'}=k_{p}'(\omega_{p+}'-\omega_{p-}')$, $\Delta k'_{s'}=k_{s}'(\omega_{s+}'-\omega_{s-}')$. Now integrating over $\omega_{s+}'$ and $\omega_{s-}$ for the first two terms gives zero, as they cancel out each other as well. Similarly by changing variables and integrating other terms will have the terms cancel out each other. Hence the interaction Hamiltonian with a dominating spectral effect from the dispersion of the medium, commutes at different times.

\end{document}